\documentclass[smallcondensed]{svjour3}
\usepackage[utf8]{inputenc}
\usepackage{graphicx}
\usepackage{amsmath}
\usepackage{amssymb}
\usepackage{natbib}
\usepackage{hyperref}
\usepackage{footmisc}
\usepackage{natbib}
\setcitestyle{sort&compress,square,comma,numbers}

\journalname{Hyperfine Interactions}

\title{Direct decay-energy measurement as a route to the neutrino mass}

\begin{document}

    \author{J. Karthein \thanks{This article contains data from the Ph.D thesis work of Jonas Karthein, enrolled at Heidelberg University, Germany. Corresponding author: \href{mailto:jonas.karthein@cern.ch}{jonas.karthein@cern.ch}} \and
            D.$\>$Atanasov\footnote{Current address: KU Leuven, Instituut voor Kern- \& Stralingsfysica, 3001 Leuven, Belgium\label{ftn:Leuven}} \and
            K.$\>$Blaum \and
            S.$\>$Eliseev \and
            P.$\>$Filianin \and
            D.$\>$Lunney \and
            V.$\>$Manea\footref{ftn:Leuven} \and
            M.$\>$Mougeot\footnote{Current address: Max-Planck-Institut f\"ur Kernphysik, 69117 Heidelberg, Germany} \and
            D.$\>$Neidherr \and
            Y.$\>$Novikov \and
            L.$\>$Schweikhard \and
            A.$\>$Welker \and
            F.$\>$Wienholtz \and
            K.$\>$Zuber}
    
    \institute{ J. Karthein \and V. Manea \and A. Welker \and F. Wienholtz
                    \at CERN, Route de Meyrin, 1211 Gen\`eve, Switzerland
            \and
                J. Karthein \and K. Blaum \and S. Eliseev \and P. Filianin
                    \at Max-Planck-Institut f\"ur Kernphysik, 69117 Heidelberg, Germany
            \and
                D. Atanasov \and A. Welker \and K. Zuber
                    \at Technische Universit\"at Dresden, 01062 Dresden, Germany
            \and
                D. Lunney \and M. Mougeot
                    \at CSNSM-IN2P3-CNRS, Universit\'e  Paris-Sud, 91400 Orsay, France
            \and
                D. Neidherr
                    \at GSI  Helmholtzzentrum  f\"ur  Schwerionenforschung, 64291 Darmstadt, Germany
            \and
                Y. Novikov
                    \at Department of Physics, St Petersburg State University, St Petersburg 198504, Russia
            \and
                Y. Novikov
                    \at Petersburg Nuclear Physics Institute, 188300 St Petersburg, Russia
            \and
                L. Schweikhard \and F. Wienholtz
                    \at Physikalisches Institut, Universit\"at Greifswald, 17489 Greifswald, Germany
                }
    \date{Received: January 29, 2019 / Accepted:  May 14, 2019}
    
    \maketitle

    \begin{abstract}
        A high-precision measurement of the $^{131}$Cs$ \rightarrow ^{131}$Xe ground-to-ground-state electron-capture $Q_{\textrm{EC}}$-value was performed using the ISOLTRAP mass spectrometer at ISOLDE/CERN. The novel PI-ICR technique allowed to reach a relative mass precision $\delta m/m$ of $1.4\cdot10^{-9}$. A mass resolving power $m/\Delta m$ exceeding $1\cdot10^7$ was obtained in only $1\,$s trapping time. Allowed electron-capture transitions with sub-keV or lower decay energies are of high interest for the direct determination of the $\nu_e$ mass. The new measurement improves the uncertainty on the ground-to-ground-state $Q_{\textrm{EC}}$-value by a factor 25 precluding the $^{131}$Cs$ \rightarrow ^{131}$Xe pair as a feasible candidate for the direct determination of the $\nu_e$ mass.
        
    \end{abstract}
    
    \newpage
    \section{Introduction}
        The determination of the neutrino rest mass is of broad interest not only in nuclear physics but also in the fields of particle and astrophysics. On the most fundamental level, the existence of a non-zero neutrino mass is not explained by the standard model. However, abundant experimental evidence by the observation of neutrino oscillations has been found in the last decades, which requires a neutrino mass and mixing. Hence, a detailed study of different neutrino properties and interactions evolved as a powerful tool in the search for the fundamental theory beyond the standard model. \cite{Bellini2014, Giunti2015} A very feasible approach for the determination of the electron-neutrino mass lies in the investigation of electron-capture (EC) reactions with energies of a few keV or lower. Here, the only particle emitted is the neutrino itself. Therefore, the smaller the decay energy of these transitions, the higher the sensitivity to the neutrino rest mass. Such transitions are found in allowed EC-transitions to excited nuclear states in the daughter nucleus.\\
        
        \noindent Electron and nuclear excitation energies are typically known to sub-keV precision. Unfortunately, the ground state masses of the decay pairs are, in most cases,  known with uncertainties well above 1$\,$keV and thus constitute the main contribution to the uncertainty of decay energies. Presently, only Penning-trap mass spectrometry (PTMS) is capable of providing mass measurements with sub-keV uncertainties. In recent years, a combination of PTMS and cryogenic microcalorimetry (MMC) \cite{Ranitzsch2012} has proven to be a very successful combination for investigating the $\beta^-$-decay in $^{187}$Re and the electron capture in $^{163}$Ho \cite{Eliseev2015}. Several other transitions have been subsequently suggested as possible candidates for neutrino physics research - the electron-capture of $^{131}$Cs to the $E^*=364.490(4)\,$keV \cite{Khazov2006} excited state in $^{131}$Xe being one example.\\
    
    \section{Experiment and analysis}
    The measurement was performed with the high-precision Penning-trap mass spectrometer ISOLTRAP \cite{Lunney2017, Kreim2013, Mukherjee2008} located at CERN's radioactive ion beam facility ISOLDE \cite{Borge2018}. There, isotopes are produced in nuclear reactions in a thick target, induced by a 1.4$\,$GeV proton beam. In the present case a uranium-carbide target was used. After surface ionization, the beam was accelerated to 50$\,$keV, magnetically separated for the ion of interest in ISOLDE's HRS separator and transported to the ISOLTRAP setup.\\
    
    \begin{figure}[ht!]
        \centering
        \includegraphics[width=\linewidth]{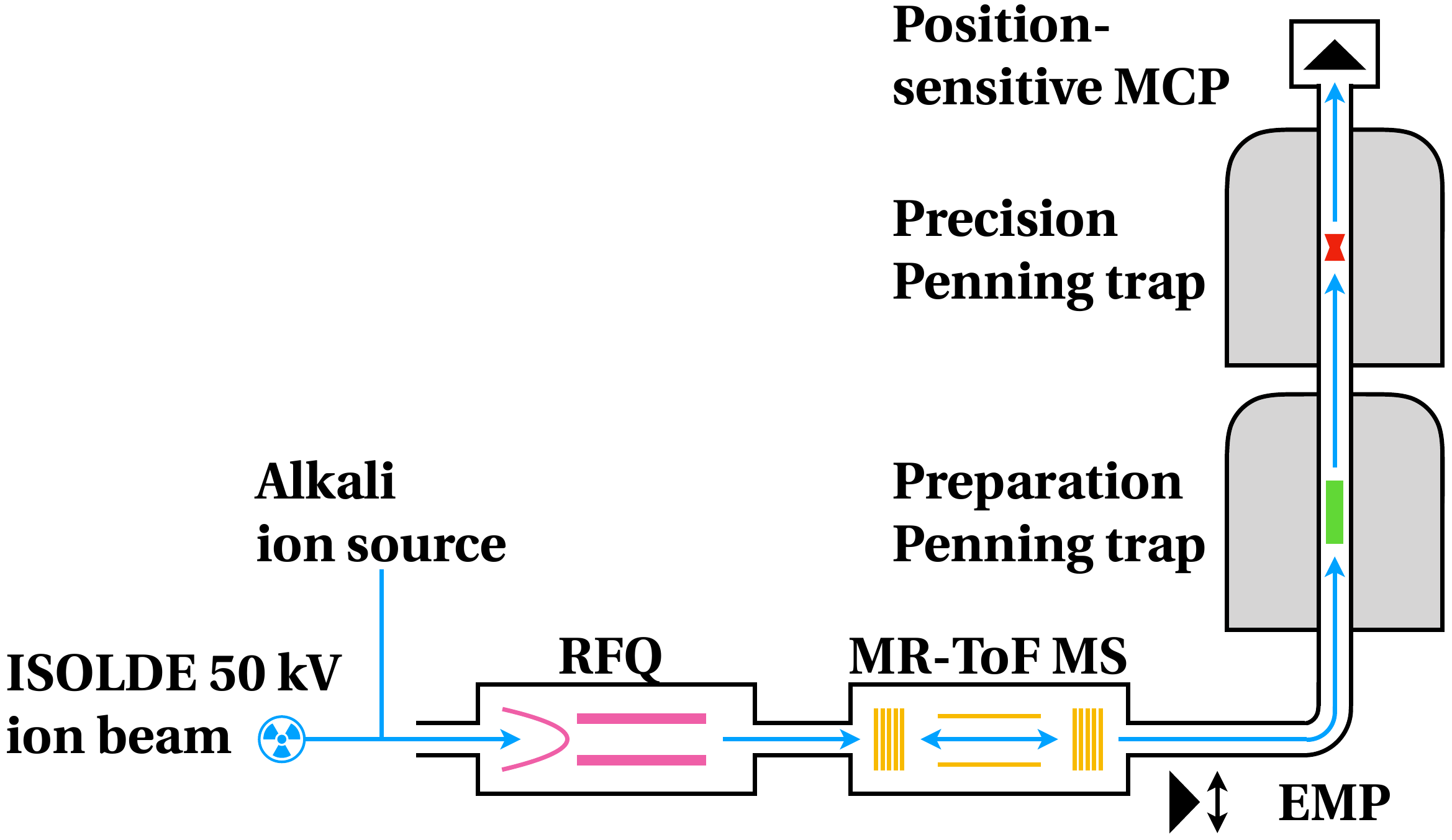}
        \caption{Schematic overview of the ISOLTRAP mass spectrometer. Radioactive ion beams provided by ISOLDE or an offline alkali ion source at an energy of 50$\,$keV are delivered. Inside the ISOLTRAP apparatus the beam is processed by a sequence of traps: a radio-frequency quadrupole (RFQ) cooler and buncher (pink), a multi-reflection time-of-flight (MR-ToF) mass separator/spectrometer (yellow), a preparation Penning trap (green) and a precision Penning trap (red). Furthermore, an electron multiplier (EMP) particle detector for ToF detection and a position-sensitive multi-channel plate (MCP) particle detector for position and ToF detection are shown. For further details, see text.}
        \label{fig:setup}
    \end{figure}
    
    \noindent The ISOLTRAP apparatus, depicted in Fig.$\>$\ref{fig:setup}, consists of a sequence of four ion traps. The continuous $^{131}$Cs$^+$ beam from ISOLDE, as well as the $^{133}$Cs$^+$ beam from ISOLTRAP's offline alkali ion source in the case of reference mass, is first accumulated in a radio-frequency quadrupole (RFQ) trap \cite{Herfurth2001}, where it is cooled and bunched for 10$\,$ms using ultra-pure helium gas. Isobaric separation is subsequently performed using ISOLTRAP's multi-reflection time-of-flight (MR-ToF) device \cite{Wolf2013}, in which trapped ions are reflected back and forth in order to extend their flight path to $\sim1\,$km ($\sim28\,$ms). Not only has this device shown numerous times its suitability for the measurement of short-lived isotopes produced in minutes quantities \cite{Mougeot2018,Wienholtz2013} but it has also proved itself to be a perfectly suitable tool for mass purification \cite{Wienholtz2017}. More specifically, in this experiment a mass resolving power $R=m/\Delta m=t/(2\cdot \Delta t)$ (where $t$ is the mean of the time-of-flight distribution and $\Delta t$ its full width at half maximum) in excess of $1.1\cdot10^5$ was achieved. The purified beam is then transported to the helium buffer-gas-filled preparation Penning trap for further cooling and purification following the well-established mass-selective centering technique \cite{Savard1991}. Ultimately, the ions arrive in the precision Penning trap where high-precision mass determination is accomplished by measuring the ion's cyclotron frequency $\nu_c$
    \begin{equation}
    \nu_c = \frac{1}{2\pi} \cdot \frac{q_i}{m_i} \cdot B
    \label{eq:cyc-freq}
    \end{equation}
    \noindent with the charge-to-mass ratio $q_i/m_i$ and the magnetic field strength $B$. All detection techniques currently available at the ISOLTRAP setup - namely the single pulse time-of-flight ion-cyclotron-resonance (ToF-ICR) mass spectrometry (MS) \cite{Konig1995}, the two-pulse Ramsey-type ToF-ICR MS \cite{George2007} and the recently developed phase-imaging ion-cyclotron-resonance (PI-ICR) MS \cite{Eliseev2013, Eliseev2014} - were all used in the presented experiment.\\
    
    \begin{figure}[ht!]
        \centering
        \includegraphics[width=\linewidth]{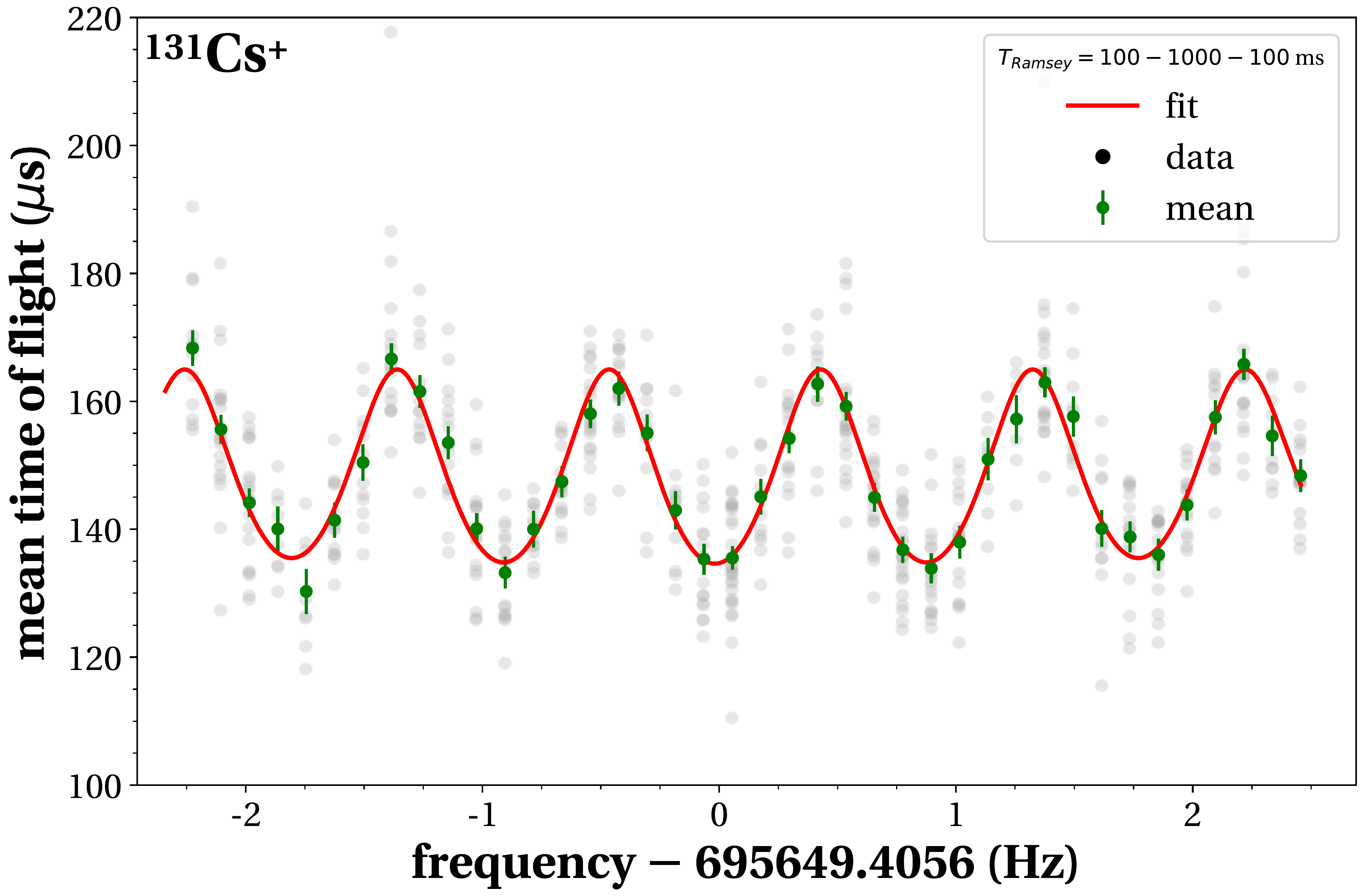}
        \caption{Typical Ramsey-type ToF-ICR spectrum of $^{131}$Cs$^+$ with an excitation time of 100$\,$ms per pulse and 1000$\,$ms waiting time. Individual, repeated ToF measurements are shown in black without any analysis cuts, thus demonstrating the purity of the beam injected inside the precision Penning trap. The mean of the unbinned ToF distribution per scan step with its standard deviation as error bar and the fitted theoretical line shape are represented in green and red respectively \cite{Konig1995}. For further details, see text.}
        \label{fig:ramsey}
    \end{figure}
    
    \noindent In both ToF-ICR techniques an excitation frequency is scanned, i.e. the excitation frequency is varied from one experimental cycle to the next, and the ion's time of flight (ToF) to a multi-channel plate detector after ejection from the trap is measured. This ToF has a minimum at the cyclotron frequency. A typical Ramsey-type ToF-ICR scan for $^{131}$Cs$^+$ is shown in Fig.$\>$\ref{fig:ramsey} for an excitation time of 100$\,$ms per pulse and a "waiting time" of 1000$\,$ms between the pulses. There, the individual, repeated ToF measurements per scan step is shown in black. The green data points represent the mean of the unbinned ToF distribution per scan step with its standard deviation as error bar. The red line represents a least squares fit of the theoretical line shape to the mean ToF distributions \cite{Konig1995}.\\

    \begin{figure}[ht!]
        \centering
        \includegraphics[width=0.6\linewidth]{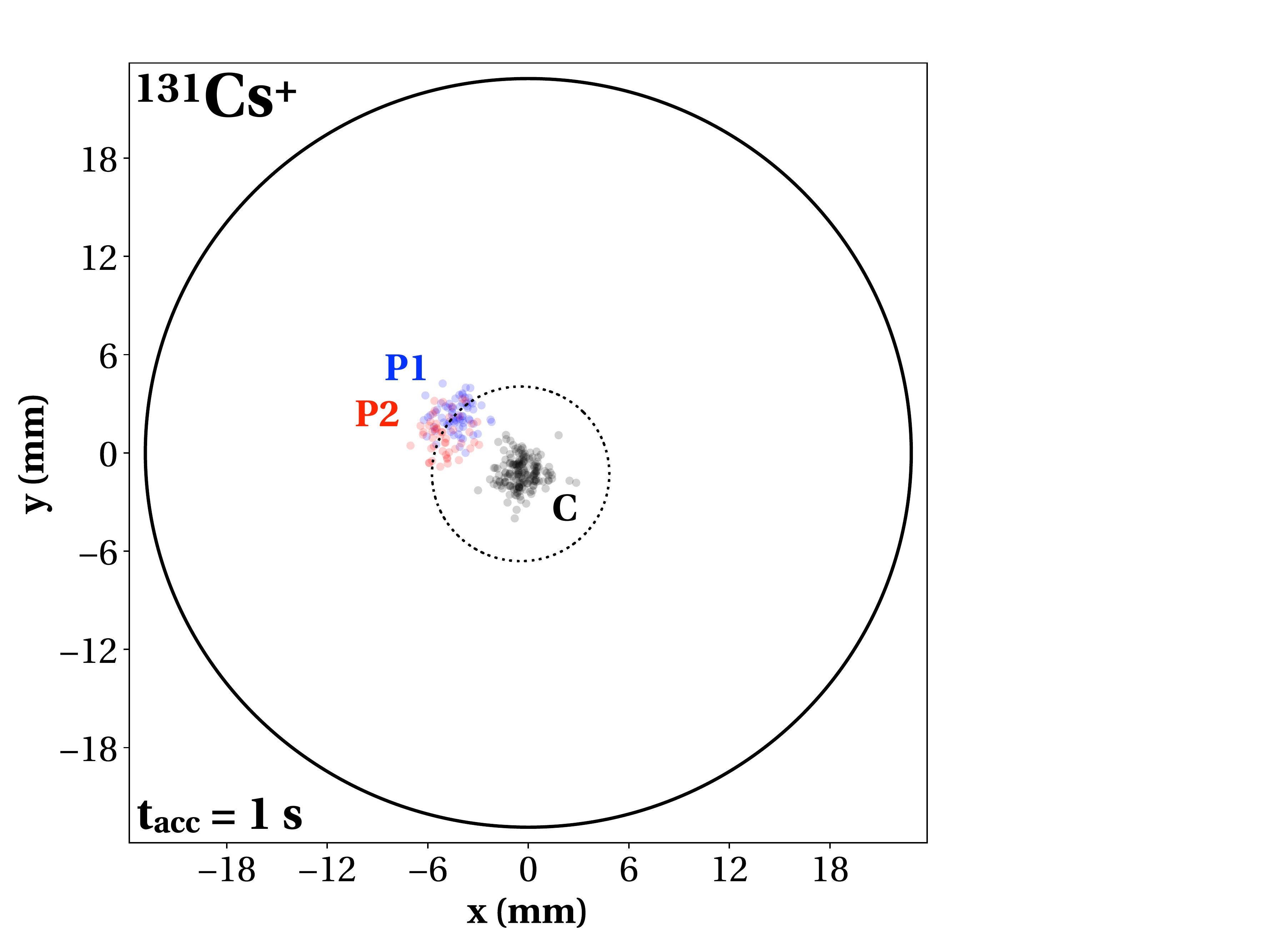}
        \caption{Typical PI-ICR detector image for $^{131}$Cs$^+$ with a center spot \textbf{C} and two overlapping spots \textbf{P1/P2} for $t_{\textrm{acc}}=1\,$s, resolving power $R=\phi_{\textrm{tot}}/(2\cdot\Delta\phi)=1\cdot10^7$. For further details, see text and Ref.$\>$\cite{Eliseev2014}.}
        \label{fig:piicr}
    \end{figure}

    \noindent In addition to the well-established ToF-ICR techniques, the new non-scanning approach to PTMS, namely PI-ICR, has been applied. This method allows the determination of radial ion frequencies by determining the full phase $\phi_{\textrm{tot}}=2\pi n+\phi$ in a given accumulation time $t_{\textrm{acc}}$, consisting of an integer number $n\in \mathbb{N}_0$ of full turns plus an additional phase $\phi$ which is measured. The radial frequency then results as $\nu_i = (2\pi n+\phi)/(2\pi t_{\textrm{acc}})$. Since the cyclotron frequency in a Penning trap $\nu_c=\nu_+ + \nu_-$ is equal to the sum of its radial eigenfrequencies $\nu_{+/-}$, the technique is perfectly suited for PTMS allowing a frequency determination at the same or better precision as ToF-ICR techniques with $\sim25$ times shorter measurement time \cite{Eliseev2014}.\\
    
    \noindent A typical PI-ICR detector image for $^{131}$Cs$^+$ is shown in Fig.$\>$\ref{fig:ramsey}: The dots represent repeated position projections (so called spots) from the Penning trap to a position-sensitive detector. In this case, the frequency determination was performed according to the pattern 1/2 (in Fig.$\>$\ref{fig:piicr} referred to as \textbf{P1/P2}) measurement scheme described in \cite{Eliseev2014} which allows for a direct determination of $\nu_c$. The achieved resolving power $R$ in case of Fig.$\>$\ref{fig:piicr} was $R=\phi_{\textrm{tot}}/(2\cdot\Delta\phi)=1\cdot10^7$ with the total accumulated phase $\phi_{\textrm{tot}}$ after $t_{\textrm{acc}}=1\,$s and the spot's FWHM in terms of angle $\Delta\phi$. The analysis was performed with a custom-designed analysis software (for details see Ref.$\>$\cite{Karthein2017a}) based on Python and ROOT \cite{Antcheva2009}. The analysis was independently performed with a LabView analysis software developped by the SHIPTRAP collaboration \cite{Eliseev2014} and agrees within uncertainties.\\

    \begin{figure}[ht!]
        \centering
        \includegraphics[width=\linewidth]{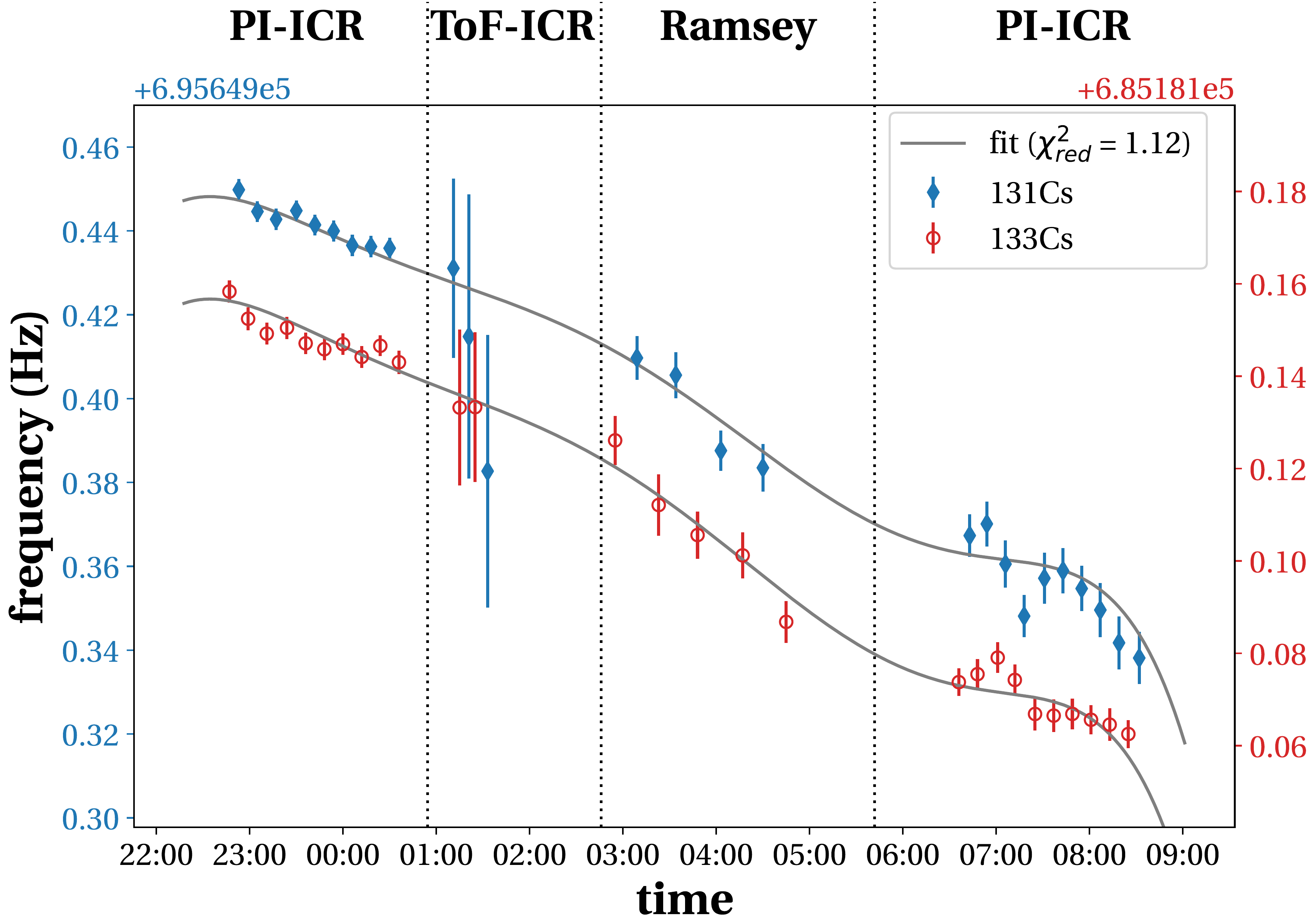}
        \caption{Simultaneous polynomial fits of the four data sets as well as of all cyclotron frequency data for $^{131}$Cs$^+$ and $^{133}$Cs$^+$. For further details, see text.}
        \label{fig:poly}
    \end{figure}
    
    \noindent The determination of the cyclotron frequency ratio $r=\nu_{\textrm{c,ioi}}/\nu_{\textrm{c,ref}}$ between all measured cyclotron frequency values $\nu_{c\textrm{,ioi}}$ of the ion of interest (in this case $^{131}$Cs$^+$) of all three measurement methods and the reference ion values $\nu_{c\textrm{,ref}}$ (in this case $^{133}$Cs$^+$) is performed by simultaneously fitting a polynomial function $p(t)$ to both data sets \cite{Fink2012}:
    \begin{eqnarray}
        \nu_{\textrm{c,ioi}} & = & p(t)\\
        \nu_{\textrm{c,ref}} & = & r\cdot \nu_{\textrm{c,ioi}} = r\cdot p(t).
    \end{eqnarray}
    \noindent The polynomial fit function describes the temporal evolution of the cyclotron frequencies while the proportionality between the two fits is exactly the cyclotron frequency ratio $r$. The ground-to-ground state $Q_\textrm{EC}$-value can be directly expressed following the relation:
    \begin{equation}
        Q_\textrm{EC}=(r-1)\cdot(m_{\textrm{ref,lit}}-m_e),
    \end{equation}
    where $m_{\textrm{ref,lit}}$ is the literature mass of the reference ion (here taken from AME16 \cite{Wang2017a}) and $m_e$ \cite{Sturm2014} is the electron mass. Figure \ref{fig:poly} shows all individual cyclotron frequency measurements of $^{131}$Cs$^+$ and $^{133}$Cs$^+$ over time. In addition, the polynomial fits are shown. As one can see, all PTMS detection methods used in this publication are in good agreement. Moreover, the weighted mean of all individual cyclotron frequency ratios for neighboring, alternating frequency measurements of $^{131}$Cs$^+$ and $^{133}$Cs$^+$ was calculated \cite{Kellerbauer2003} and agrees with the polynomial method described above. The final frequency ratio yields $r_{\textrm{final}}=0.9849517704(14)$. The uncertainty of the combination of all PI-ICR data is $\delta\nu_c/\nu_c = 1.4\cdot10^{-9}$.\\

    \noindent In addition to the statistical uncertainty derived from the fit, a careful analysis of the systematic uncertainties which are not covered by the polynomial fit was performed. These include considering fit parameter correlations, where off-axis elements in the correlation matrix were negligibly small. The fluctuation of the individual frequencies after applying different fit cuts was systematically studied. They were found to be well within the statistical uncertainty on the individual frequency, proofing the purity of the beam. Since the ion rate was purposely kept below one ion per measurement cycle, a z-class analysis, i.e. reducing the number of detected ions per cycle and therefore in the trap itself, did not have to be performed. The data was corrected for ISOLTRAP's mass-dependent shift (relative shift: $7\cdot10^{-10}$) due to the difference in mass between the ion of interest and the reference ion as described in Ref.$\>$\cite{Kellerbauer2003, Welker2017ab}. The residual systematic uncertainty of ISOLTRAP \cite{Kellerbauer2003} was not taken into account due to the fact that both the ion of interest and the reference were prepared, injected and measured in identical conditions, hence probing the same volume of the precision trap.\\
    
    \noindent Table $\>$\ref{tab:results} presents the obtained ground-to-ground-state decay energy $Q_\textrm{EC}$ as well as the allowed ($Q_\textrm{EC}-E^*$)-value of interest to the $E^*=364.490(4)\,$keV \cite{Khazov2006} state in $^{131}$Xe with their associated uncertainties. The decay-energy of the allowed EC-transition has to be corrected for the binding energy $B$ of captured electrons ($Q_\textrm{EC}-E^*-B$) to the $L-$shell-electron ($B(L$-$e^-)=5.453\,$keV  \cite{Cardona1978, NIST}) and to the $M-$shell-electron ($B(M$-$e^-)=1.1487\,$keV  \cite{Cardona1978, NIST}). It is worth mentioning, that the $^{131}$Xe literature mass is dominated by a high-precision measurement from SHIPTRAP using the PI-ICR technique \cite{Nesterenko2014}.\\
    
    \begin{table}[t!]
    \centering
    \caption{Comparison of the measured mass excess, the measured released energy $Q_\textrm{EC}$ of the electron-capture-pair ground-to-ground-state decay of $^{131}$Cs$ \rightarrow ^{131}$Xe, the released energy ($Q_\textrm{EC}-E^*$) for this electron-capture-pair in terms of ground-to-excited-state decay of $^{131}$Cs$ \rightarrow ^{131}$Xe$^*$, the latter one corrected for the binding energy $B$ of captured $L$- and $M$-shell electrons ($Q_\textrm{EC}-E^*-B$) and the final uncertainty compared to literature \cite{Khazov2006, Wang2017a}. For further details, see text.}
        \label{tab:results}
        \begin{tabular}{ccccccc}
        \hline\noalign{\smallskip}
        (keV) & $ME$ & $Q_{\textrm{EC}}$ & $Q_\textrm{EC}$-$E^*$ &  $Q_\textrm{EC}$-$E^*$-$B_L$ & $Q_\textrm{EC}$-$E^*$-$B_M$ & unc.\\
        \noalign{\smallskip}\hline\noalign{\smallskip}
        Literature & -88059 & 355 & -10 & -15 & -11 & 5\\
        ISOLTRAP & -88055.56 & 358.00 & -6.49 & -11.95 & -7.64 & 0.17 \\
        \noalign{\smallskip}\hline
        \end{tabular}
    \end{table}
    
    \noindent With the refined uncertainty, the ground-to-excited-state value $(Q_\textrm{EC}-E^*)=-6.49(17)\,$keV appears undoubtedly negative. This translates to the excited state $^{131}$Xe$^*$ being higher in energy than the parent ground state in  $^{131}$Cs, thus prohibiting this $^{131}$Cs$ \rightarrow ^{131}$Xe$^*$ transition and excluding it as a suitable candidate for the determination of the electron-neutrino mass.

    \section{Conclusion}
        High-precision mass measurements of the $^{131}$Cs using established time-of-flight ion-cyclotron-resonance (ToF-ICR) mass spectrometry (MS) as well as the recently developed phase-imaging ion-cyclotron-resonance (PI-ICR) detection technique was performed with ISOLTRAP/CERN. We were able to demonstrate the successful implementation of PI-ICR at ISOLTRAP with a high resolving power of $1\cdot10^7$ for 1$\,$s single-measurement time, a statistical uncertainty of only $1.4\cdot10^{-9}$ in $\sim4\,$hrs of beam time and a very good agreement with our well-established ToF-ICR measurement techniques (see Fig.$\>$\ref{fig:poly}). The obtained $Q_{\textrm{EC}}$-value agrees with the value found in literature. However, the refined precision allows now to exclude this electron-capture transition as a possible candidate for the determination of the neutrino mass.\\
        
        \noindent Thus the PI-ICR technique appears very promising to tackle even more challenging cases such as $^{134}$Ce, $^{159}$Dy and $^{175}$Hf \cite{Eliseeva, Karthein2017}, the decay energy of which must be determined at a sub-100 eV level of precision.
    
    \begin{acknowledgements}
        We thank the ISOLDE technical group and the ISOLDE Collaboration for their professional help. We acknowledge support by the Max Planck Society, the German Federal Ministry of Education and Research (BMBF) (05P12HGCI1, 05P12HGFNE, and 05P15ODCIA), the French IN2P3, the ExtreMe Matter Institute (EMMI) at GSI, and the European Union’s Horizon 2020 research and innovation programme (654002). Jonas Karthein acknowledges the support by a Wolfgang Gentner Ph.D scholarship of the BMBF (05E12CHA).
    \end{acknowledgements}

    
\end{document}